\documentclass[fleqn,twoside]{article}
\usepackage{espcrc2}
\usepackage{graphics}
\def\lsim{\raise0.3ex\hbox{$<$\kern-0.75em\raise-1.1ex\hbox{$\sim$}}}
\def\gsim{\raise0.3ex\hbox{$>$\kern-0.75em\raise-1.1ex\hbox{$\sim$}}}

\newcommand{\pslash}{p\kern-1ex /}
\newcommand{\Dslash}{{\cal D}\kern-1.5ex /}

\newcommand{\beqa}{\begin{eqnarray}}
\newcommand{\eeqa}{\end{eqnarray}}

\newcommand{\beq}{\begin{equation}}
\newcommand{\eeq}{\end{equation}}
\newcommand{\bc}{\begin{center}}
\newcommand{\ec}{\end{center}}
\title{
 Neutron Electric Dipole Moment with Domain Wall Quarks
\author{
F. Berruto
\address[BNL]{Physics Department, Brookhaven National Laboratory, Upton, NY 11973 USA},
T. Blum
\address{Physics Department, University of Connecticut, Storrs, 
CT 06269-3046 and \\~~RIKEN BNL Research Center, Brookhaven National Laboratory, 
Upton, NY 11973 USA},
 K. Orginos
\address[MIT]{Center for Theoretical Phyics, Massachusetts Institute of Technology, Cambridge, MA, 02139 },
A. Soni$~^a$
}}
\begin{document}
\pagestyle{empty}

\begin{abstract}
We present preliminary results for nucleon dipole moments computed with domain wall fermions. Our main target is the electric dipole moment of the neutron arising from the $\theta$ term in the gauge part of the QCD  lagrangian. The calculated magnetic dipole moments of the proton and neutron are in rough accord with experimental values.
\end{abstract}

\maketitle

\section{Introduction}
\label{sec:intro}

One of the most intriguing aspects of QCD is that it allows a T and
P-odd,
gauge invariant interaction term, the so called $\theta$ term.
The presence of such a term has the profound effect that the
Strong interactions violate CP. One of the best ways to monitor
this is by searching for the neutron electric dipole moment,
$d_N$.
In the Standard Model, the CP-odd phase of the CKM mixing matrix can also
trigger a non-vanishing value for $d_N$, but this can not
occur at one loop order (in the Weak interaction)
and is consequently estimated to
be $\le 10^{-30}\, e$-cm, many orders of magnitude
below the current experimental bound\cite{expts},
$d_N = |\vec{d}_N| < 6.3 \times 10^{-26}\,e$-cm.
Using this experimental bound along with various
estimates of $d_N$ due to the $\theta$ term
then invariably implies
that the CP-odd parameter in the QCD action,
$\theta \le 10^{-10}$,
is exceedingly and indeed {\it unnaturally} small. Since
there is no good reason for this number to be so
different from unity, its minuteness requires extraordinary
fine-tuning.  This is
often termed the {\it Strong CP problem}.

To translate the above experimental bound to a constraint
on the fundamental $\theta$ parameter requires
evaluation of nucleon matrix elements, wherein lattice
QCD enters.

There have been several past attempts to calculate $d_N$,
in the continuum and on the lattice.
\begin{enumerate} 

\item
~V.~Baluni~\cite{baluni}
%[``CP Violating Effects In QCD,''\cite{baluni}
%Phys.\ Rev.\ D {\bf 19}, 2227 (1979)]
computed $d_N$ in the framework of
the
MIT bag model obtaining 
$
d_N\simeq 8.2\cdot 10^{-16}\theta e\cdot cm
$

\item

Crewther {\it et al.}
\cite{witten},
using an effective chiral lagrangian found 
$
d_N\propto \theta \  M_{\pi}^2\ \mbox{ln}(M_{\pi}^2)\simeq
5.2\cdot 10^{-16}\theta\, e$-cm
\item
Pospelov and Ritz\cite{Pospelov:1999ha}, using QCD sum rules techniques, found
$
d_N = 1.2 \times 10^{-16}\theta\, e$-cm
\item

Aoki and Gocksch\cite{sinya1}
%(Phys.Rev.Lett.63:1125,1989)
were the first
to try a pioneering  lattice calculation of the $d_N$ in the quenched
approximation.

\end{enumerate}

A non-zero $d_N$ is induced by the $\theta$-term in the QCD gauge action.
\begin{eqnarray}
S_{\theta}&=&\int d^4x\  i\theta \frac{g^2}{32\pi^2}\mbox{tr}\left[G(x)
\tilde{G}(x)\right] = i \theta Q.
 %\label{stheta}
 \end{eqnarray}
where $Q$ is the topological charge of the QCD vacuum.
Note the factor of $i$ in the $\theta$ term makes the action complex
and therefore difficult to handle in lattice simulations for arbitrary $\theta$. If $\theta$ is small, as discussed below, the difficulty can
be avoided.

 Using the axial anomaly, one can replace the CP violating gauge
 action above with the fermionic action,
 $
 S'_{\theta}=-i\theta\overline{m}\int d^4x P(x)
 $
 where
$
 P(x)=\bar{u}(x)\gamma_5 u(x)+
 \bar{d}(x)\gamma_5 d(x)+
 \bar{s}(x)\gamma_5 s(x)
 $
 and $\overline{m}^{-1}=m_u^{-1}+m_d^{-1}+m_s^{-1}$. 
 In \cite{sinya1} $d_N$ was evaluated by
extracting the spin-up and spin-down neutron masses from the two-point functions obtained by adding to the action $S'_{\theta}$ and a term corresponding to a constant background electric field oriented in a fixed spatial direction. 
Since $\theta$ is small, it is sufficient to consider a
single insertion of $S'_{\theta}$ in the 2-point functions.

Since the operator insertion in diagram (a) of \cite{sinya1}
is manifestly CP-conserving, its contribution to a CP-odd physical observable
(such as electric dipole moment) must vanish. Contributions to
$d_N$ therefore must come  from diagram
(b) of \cite{sinya1} which was ignored in that calculation.

More detailed reasoning for the fact that $d_N$ resides only in
fig(b) of \cite{sinya1} was given shortly thereafter by \cite{sinya2};
a significant extension of that analysis appeared recently in \cite{marti}.
As stressed in \cite{sinya2}, the quenched approximation does not
justify ignoring diagram (b) of \cite{sinya1}.

\section{ Our computational strategy}

There are two important elements to our computational strategy:

(1) Compute the matrix elements of the electromagnetic current between nucleon states,
$\langle p^\prime,s| J^\mu | p,s\rangle_\theta
=\bar u(p^\prime,s)\Gamma_\mu(q^2)u(p,s)$, where
\begin{eqnarray}
\Gamma_\mu(q^2) &=&
\gamma_\mu\,F_1(q^2)
+i\,\sigma_{\mu\nu}q^\nu\,\frac{F_2(q^2)}{2m}\\\nonumber
&&+\left(\gamma_\mu\,\gamma_5\,q^2\,
          - 2m\gamma_5\,q_\mu\right)F_A(q^2)\\\nonumber
&&+\sigma_{\mu\nu}q^\nu\gamma_5\,\frac{F_3(q^2)}{2m},
\end{eqnarray}
and use projectors to obtain linear combinations of $F_1$ and $F_2$,
\begin{eqnarray}
G_E(q^2) &=&F_1(q^2) + \frac{q^2}{(2m)^2} F_2(q^2)\\
G_M(q^2)&=& F_1(q^2)+F_2(q^2),
\end{eqnarray}
and $F_3(q^2)/2m$. By forming ratios of $G_M(q^2)$ and $F_3(q^2)$ with $G_E(q^2)$ and taking $q^2\to0$, we find both 
magnetic and electric dipole moments, respectively: $e(F_1(0)+a_\mu)/2m$ and $d_N= e F_3(0)/2m$
(note, $G_E(0)=F_1(0)=1$ and 0 for the proton and neutron, respectively, and $a_\mu\equiv F_2(0)$ is the anomalous magnetic moment). In the above $q^2\le0$, and $m$ is the nucleon mass.

(2) Expand $\langle p^\prime,s| J^\mu | p,s \rangle_\theta$ to lowest order in $\theta$ and compute $F_3(q^2)$
in each topological sector $\nu$, and then average over all sectors
with weight $Q_\nu$.

A disadvantage of this method is
that unlike the background-electric-field-method\cite{cwb}
used in \cite{sinya1}, our method does not allow
a direct calculation of
the electric dipole moment, {\it i.e.} the value of the
form-factor at $q^2=0$ since 
on a finite lattice only the form factor $F_1$ in Eq. 2
can be computed
at $q^2=0$ \cite{wilcox}.  
Our method
requires extrapolation of the form factors 
to $q^2=0$ from non-vanishing values of $q^2$. 

\section{Remarks on the quenched case}

The QCD partition function in the presence of explicit CP violation is
$
Z(\overline{\eta},\eta)=
\int {\cal D}A_{\mu}\ \mbox{det}[D(m)+i\theta \overline{m}\gamma_5]
\ e^{\overline{\eta} D(m)^{-1}\eta-S_G}.
$
Setting $\mbox{det}[D(m)+i\theta \overline{m}\gamma_5]=1$, we lose
CP violating physics. However, if $\theta$ is small,
$\det{[D(m)+i{ \theta \overline{m}\gamma_5}]} =
\mbox{det}[D(m)] \ [1+i{\theta \overline{m}\
\mbox{tr}(\gamma_5 D
(m)^{-1})}\ ] + {\cal O}(\theta^2),
$
and we quench as usual by setting $\det{[D(m)]}=1$. Considering disconnected insertions
of $P=\mbox{tr}(\gamma_5 D(m)^{-1})$,
it seems clear that one can compute the disconnected diagram
(b) of \cite{sinya1} 
using quenched gauge configurations and obtain a non-zero result.

\section{The chiral limit}

%\begin{eqnarray}
%D(0)|\lambda_i\rangle &=& \lambda_i |\lambda_i\rangle \nonumber\\
%D(m)|\lambda_i\rangle &=& (D(0)+m) |\lambda_i\rangle =
%(\lambda_i+m) |\lambda_i\rangle \nonumber
%\end{eqnarray}
The spectral decomposition of $D^{-1}(m)$
leads to 
\begin{eqnarray}
\sum_{f=1}^{N_f}\mbox{Tr}\left[\gamma_5
D^{-1}(m_f)\right]=\frac{n_+-n_-}
{\overline{m}}=\frac{Q}{\overline{m}}
\end{eqnarray}
for $N_f$ flavors and $n_+$ and $n_-$ the number of right- and left-handed zero modes of $D(m)$. 
If we trade $Q$ for a disconnected insertion of $-\overline{m}P$, $d_N$ will vanish in the chiral limit only if $(\det{D(m)})^{N_f}$ vanishes, and thus $d_N$ can not vanish in the quenched chiral limit (recall that $\det{ D(m)} \sim m$ for $Q\neq 0$, and contributions to $d_N$ vanish for $Q=0$.).

\section{ Numerical Results}

In Table 1 we summarize results for the ratios of neutron and proton magnetic to proton electric form factors which become the dipole moment in question in the limit $q^2\to 0$. These results were computed on 280  $N_f=2$, $m_{sea}=0.02$, domain wall fermion configurations (separated by 10-15 trajectories) with $m_{val}=0.04$ and 0.08\cite{rbc-dyn}. The lattice size is $16^3\times 32$, $L_s=12$, and the inverse lattice spacing in the $m_{sea}=0$ limit is $a^{-1}\approx 1.7$ GeV.
We have averaged over time slices 14-17 and (equivalent) permutations of momenta, 
$\vec p=(1,0,0)$, (1,1,0), and (1,1,1).
$Q$ was computed by integrating the topological charge
density after APE smearing the gauge fields (20 sweeps with ape weight 0.45)~\cite{DeGrand:1997ss}. 
We are also investigating
computing the topological charge from the index defined from the domain wall fermion Dirac operator (strictly valid in the limit $L_s\to \infty$).

\begin{table}[htb]
\vskip -.15in
\caption{Ratios of form factors for $m_{val}=0.04$ (upper) and 0.08
($^P$ proton, $^N$ neutron).}
\begin{center}
\begin{tabular}{|c|c|c|}
\hline
Ratio & $(a\vec{p})^2$ & value (error) \cr
\hline
$G_{M}^{P}/(E+M)/G^P_E$  & 1 & 1.363 (57)\cr
$G_{M}^{P}/(E+M)/G^P_E$ & 2 & 1.392 (61)\cr
$G_{M}^{P}/(E+M)/G^P_E$ & 3 & 1.540 (91)\cr
\hline
$G_{M}^{N}/(E+M)/G^P_E$ & 1 & -0.836 (38)\cr
$G_{M}^{N}/(E+M)/G^P_E$  & 2 & -0.837 (42)\cr
$G_{M}^{N}/(E+M)/G^P_E$ & 3 & -0.927(61)\cr
\hline
$F_{3}^{N}/(E+M)/ G^P_E$ & 1 & -0.064 (154)\cr
$F_{3}^{N}/(E+M)/ G^P_E$  & 2 & -0.037 (157)\cr
$F_{3}^{N}/(E+M)/ G^P_E$  & 3 & 0.150 (172)\cr
\hline
\hline
$G_{M}^{P}/(E+M)/G^P_E$  & 1 & 1.183 (21)\cr
$G_{M}^{P}/(E+M)/G^P_E$ & 2 & 1.194 (22)\cr
$G_{M}^{P}/(E+M)/G^P_E$ & 3 & 1.205 (27)\cr
\hline
$G_{M}^{N}/(E+M)/G^P_E$ & 1 & -0.739 (15)\cr
$G_{M}^{N}/(E+M)/G^P_E$  & 2 & -0.740 (16)\cr
$G_{M}^{N}/(E+M)/G^P_E$ & 3 & -0.747 (19)\cr
\hline
$F_{3}^{N}/(E+M)/ G^P_E$ & 1 & -0.048 (53)\cr
$F_{3}^{N}/(E+M)/ G^P_E$  & 2 & -0.029 (57)\cr
$F_{3}^{N}/(E+M)/ G^P_E$  & 3 & 0.028 (62)\cr
\hline\end{tabular}
\end{center}
\label{tab: dipole results}
\vskip -.25in
\end{table}%

Similar results hold at $m_f=0.08$, but with smaller errors. 
Approximating the $q^2\to 0$ value of each 
with the smallest value of $\vec{p}^2~(=(2\pi/16)^2)$, 
taking $m_f=0.04$ as the physical 
value of the light quark mass, and $aM_N=aM_P=0.8989(77)$ at this quark mass, we obtain
\begin{eqnarray*}
a^P_\mu &\approx& 1.45(10)\\
a^N_\mu &\approx&  -1.50(7)\\
d_N/(e\,a\,\theta) &\approx& -0.06(15).
\end{eqnarray*}
Considering the crude extrapolations just described,
these are roughly consistent with 
the experimental values $a_\mu^P = 1.79$ and $a_\mu^N=-1.91$ (and of
course $d_N\sim0$).The error estimates are statistical uncertainties only.
In physical units, $d_N=-7.4(18.0)\times 10^{-16}\,\theta\,e$-cm, consistent with the model calculations mentioned above.
We note that $a_\mu$ increases in magnitude for both the proton and neutron as $m_{val}$ decreases. The electric dipole moment of the
neutron does not show any significant dependence on the quark mass, within (relatively large) statistical errors. In the near future, we will strive to reduce the statistical error on our determination of $d_N$ which has
already yielded an interesting first-principles bound on the magnitude of $d_N$.

\vskip -.25in
\section*{Acknowledgement}
The computations described here were done on the RIKEN BNL Research Center QCDSP supercomputer. We thank our colleagues in the RBC collaboration, and in particular N. Christ and M. Creutz, for useful discussions. The work of FB and AS was supported in part by US DOE grant \# DE-AC02-98CH10886. KO was supported in part by DOE grant \#DFFC02-94ER40818. 
%TB was supported by the RIKEN BNL Research center.

%
%
%--------------------------------------------------------------------

\end{document}